\documentclass{emulateapj}
\usepackage{psfig}
\usepackage{color}
\usepackage{atbegshi}
\def\hackaltaffiltext#1#2{\AtBeginShipoutNext{\footnotetext[#1]{#2}\stepcounter{footnote}}}

\def\simlt{\lower.5ex\hbox{\ltsima}}

\newcommand{\dgr}{$^{\circ}~$}

\newcommand{\hi}{H{\footnotesize I} }

\newcommand{\midline}{\, | \,}

\begin{document}

\title{Exploring the Very Extended Low Surface Brightness Stellar \\ Populations of the Large Magellanic Cloud with SMASH}
%\title{Exploring the Very Extended Stellar Populations of \\ the Large Magellanic Cloud with SMASH}
%\title{Discovery of Very Extended Stellar Populations of \\ the Large Magellanic Cloud in the SMASH Survey}

\shorttitle{LMC Stellar Envelope}
\shortauthors{D. L. Nidever et al.}

\author{
David L. Nidever\altaffilmark{1,2},
Knut Olsen\altaffilmark{2},
Yumi Choi\altaffilmark{3},
Thomas J. L. de Boer\altaffilmark{4},
Robert D. Blum\altaffilmark{2},
Eric F. Bell\altaffilmark{5},
Dennis Zaritsky\altaffilmark{3},
Nicolas F. Martin\altaffilmark{6,7}
Abhijit Saha\altaffilmark{2},
Blair C. Conn\altaffilmark{8,9},
Gurtina Besla\altaffilmark{3},
Roeland P. van der Marel\altaffilmark{10},
Noelia E. D. No\"el\altaffilmark{4},
Antonela Monachesi\altaffilmark{11,12},
Guy S. Stringfellow\altaffilmark{13},
Pol Massana\altaffilmark{4},
Maria-Rosa L. Cioni\altaffilmark{14},
Carme Gallart\altaffilmark{15,16},
Matteo Monelli\altaffilmark{15,16},
David Martinez-Delgado\altaffilmark{17},
Ricardo R. Mu\~noz\altaffilmark{18,19},
Steven R. Majewski\altaffilmark{20},
A. Katherina Vivas\altaffilmark{21},
Alistair R. Walker\altaffilmark{21},
Catherine Kaleida\altaffilmark{10}, and
You-Hua Chu\altaffilmark{22}
%Carme Gallart\altaffilmark{9,10},
%Edward W. Olszewski\altaffilmark{2},
%Robert A. Gruendl\altaffilmark{6,7},
%You-Hua Chu\altaffilmark{11,7},
%Maria-Rosa L. Cioni\altaffilmark{13,14,15},
%Shoko Jin\altaffilmark{19},
%Andrea Kunder\altaffilmark{13},
%Matteo Monelli\altaffilmark{17,18},
%Lara Monteagudo\altaffilmark{17,18},
%Guy S. Stringfellow\altaffilmark{26},
}

\altaffiltext{1}{Department of Physics, Montana State University, P.O. Box 173840, Bozeman, MT 59717-3840 (dnidever@montana.edu)}
\altaffiltext{2}{National Optical Astronomy Observatory, 950 North Cherry Ave, Tucson, AZ 85719}
\altaffiltext{3}{Steward Observatory, University of Arizona, 933 North Cherry Avenue, Tucson AZ, 85721}
\altaffiltext{4}{Department of Physics, University of Surrey, Guildford, GU2 7XH, UK}
\altaffiltext{5}{Department of Astronomy, University of Michigan, 1085 S. University Ave., Ann Arbor, MI 48109-1107, USA}
\altaffiltext{6}{Observatoire astronomique de Strasbourg, Universit\'e de Strasbourg, CNRS, UMR 7550, 11 rue de l'Universit\'e, F-67000 Strasbourg, France}
\altaffiltext{7}{Max-Planck-Institut f\"ur Astronomie, K\"onigstuhl 17, D-69117 Heidelberg, Germany}
\altaffiltext{8}{Research School of Astronomy and Astrophysics, Australian National University, Canberra, ACT 2611, Australia}
\altaffiltext{9}{Gemini Observatory, Recinto AURA, Colina El Pino s/n, La Serena, Chile.}
\altaffiltext{10}{Space Telescope Science Institute, 3700 San Martin Drive, Baltimore, MD 21218}
\altaffiltext{11}{Instituto de Investigaci\'on Multidisciplinario en Ciencia y Tecnolog\'ia, Universidad de La Serena, Ra\'ul Bitr\'an 1305, La Serena, Chile}
\altaffiltext{12}{Departamento de F\'isica y Astronom\'ia, Universidad de La Serena, Av. Juan Cisternas 1200 N, La Serena, Chile}
\altaffiltext{13}{Center for Astrophysics and Space Astronomy, University of Colorado, 389 UCB, Boulder, CO, 80309-0389, USA}
\altaffiltext{14}{Leibniz-Institut f\"{u}r Astrophysics Potsdam (AIP), An der Sternwarte 16, 14482 Potsdam Germany}
\hackaltaffiltext{15}{Instituto de Astrof\'{i}sica de Canarias, La Laguna, Tenerife, Spain}
\hackaltaffiltext{16}{Departamento de Astrof\'{i}sica, Universidad de La Laguna, Tenerife, Spain}
\hackaltaffiltext{17}{Astronomisches Rechen-Institut, Zentrum f\"ur Astronomie der Universit\"at Heidelberg,  M{\"o}nchhofstr. 12-14, 69120 Heidelberg, Germany}
\hackaltaffiltext{18}{Departamento de Astronom\'ia, Universidad de Chile, Camino del Observatorio 1515, Las Condes, Santiago, Chile}
\hackaltaffiltext{19}{Visiting astronomer, Cerro Tololo Inter-American Observatory, National Optical Astronomy Observatory, which is operated by the Association of Universities for Research in Astronomy (AURA) under a cooperative agreement with the National Science Foundation.}
%\phantom{text text text text text text text text text text text text text text text text text text text text text text text text text text text text tex}
\hackaltaffiltext{20}{Department of Astronomy, University of Virginia, Charlottesville, VA 22904, USA}
\hackaltaffiltext{21}{Cerro Tololo Inter-American Observatory, National Optical Astronomy Observatory, Casilla 603, La Serena, Chile}
\hackaltaffiltext{22}{Institute of Astronomy and Astrophysics, Academia Sinica, P.O. Box 23-141, Taipei 10617, Taiwan, R.O.C.}

\begin{abstract}
We present the detection of very extended stellar populations around the Large Magellanic Cloud (LMC) out to R$\sim$21\degr, or $\sim$18.5 kpc at the LMC distance of 50 kpc, as detected in the Survey of the \textsc{Ma}gellanic Stellar History (SMASH) performed with the Dark Energy Camera on the NOAO Blanco
4m Telescope.  The deep ($g$$\sim$24) SMASH color magnitude diagrams (CMDs) clearly reveal old ($\sim$9 Gyr), metal-poor ([Fe/H]$\approx$$-$0.8 dex)
main-sequence stars at a distance of $\sim$50 kpc.
%well separated from the contamination of foreground Milky Way stars and unresolved background
%galaxies after our star/galaxy separation scheme is applied.
The surface brightness of these detections is extremely low with our most distant detection at $\Sigma_{g}$$\approx$34 mag arcsec$^{-2}$.
The SMASH radial density profile breaks from the inner LMC exponential decline at $\sim$13--15\dgr and a second component at larger radii has a shallower slope
with power-law index $\alpha$=$-$2.2 that contributes $\sim$0.4\% of the LMC's total stellar mass.
In addition, the SMASH densities exhibit large scatter around our best-fit model of $\sim$70\% indicating that the envelope of stellar material in the LMC
periphery is highly disturbed.
We also use data from the NOAO Source catalog to map the LMC main-sequence populations at intermediate radii and detect a steep dropoff in density
on the eastern side of the LMC (at $R$$\approx$8\degr) as well as an extended structure to the far northeast.
%indicative of a classical stellar halo or stellar tidal debris.
These combined results confirm the existence of a very extended, low-density envelope of stellar material with disturbed shape around the LMC.
The exact origin of this structure remains unclear but the leading options include a classical accreted halo or tidally stripped outer disk material.
%as presented by previous studies, which can be interpreted as either a classical accreted halo or tidally stripped material from the outer disk.
\end{abstract}

% stellar halo??

% Outline:
% Introduction
% Observations and Data Reduction
% Results
% Discussion

% Figures
% -map of SMASH fields, with updated DES footprint and fields in question marked
%    fields observed also marked?
% -many panel plot with CMDS
% -figure showing how star/galaxy separation works
% -figure showing the outmost fields with foreground subtracted with Besancon model
% -radial density profile with MAPS, OLS and SMASH data,  use a "common" field to scale things

% Introduction
\section{Introduction}
%Knut
The advent of deep wide-area surveys in the Southern hemisphere has dramatically changed our view of the size and structure of the Magellanic Clouds, particularly through the use of resolved stars as tracers of structure.  Using red giant branch (RGB) and asymptotic giant branch (AGB) stars measured in the 2MASS \citep{Skrutskie06} and DENIS \citep{Epch97} surveys, \citet{vandermarel2001} showed that the disk of the LMC appears both more extended, smoother, and more elongated than optical photographs suggest in these maps of intermediate-age and old stellar populations.  \citet{Choi2018a} extended the reach and resolution of such maps further by using red clump stars measured by the 
Survey of the MAgellanic Stellar History\footnote{\url{http://datalab.noao.edu/smash/smash.php}} \citep[SMASH;][]{Nidever2017}, and found a distinct warp in the southwestern portion of the outer disk, 7\dgr from the center, bending $\sim$4 kpc out of the LMC plane. 
In addition, \citet{Choi2018b} used the red clump maps to find a ring-like stellar overdensity in the LMC disk at a radius of $\sim$6\dgr ($\sim$5.2 kpc) with an amplitude of up to $\sim$2.5 times larger than the smooth disk. 

\begin{figure*}[ht!]
\begin{center}
\includegraphics[width=0.49\hsize,angle=0]{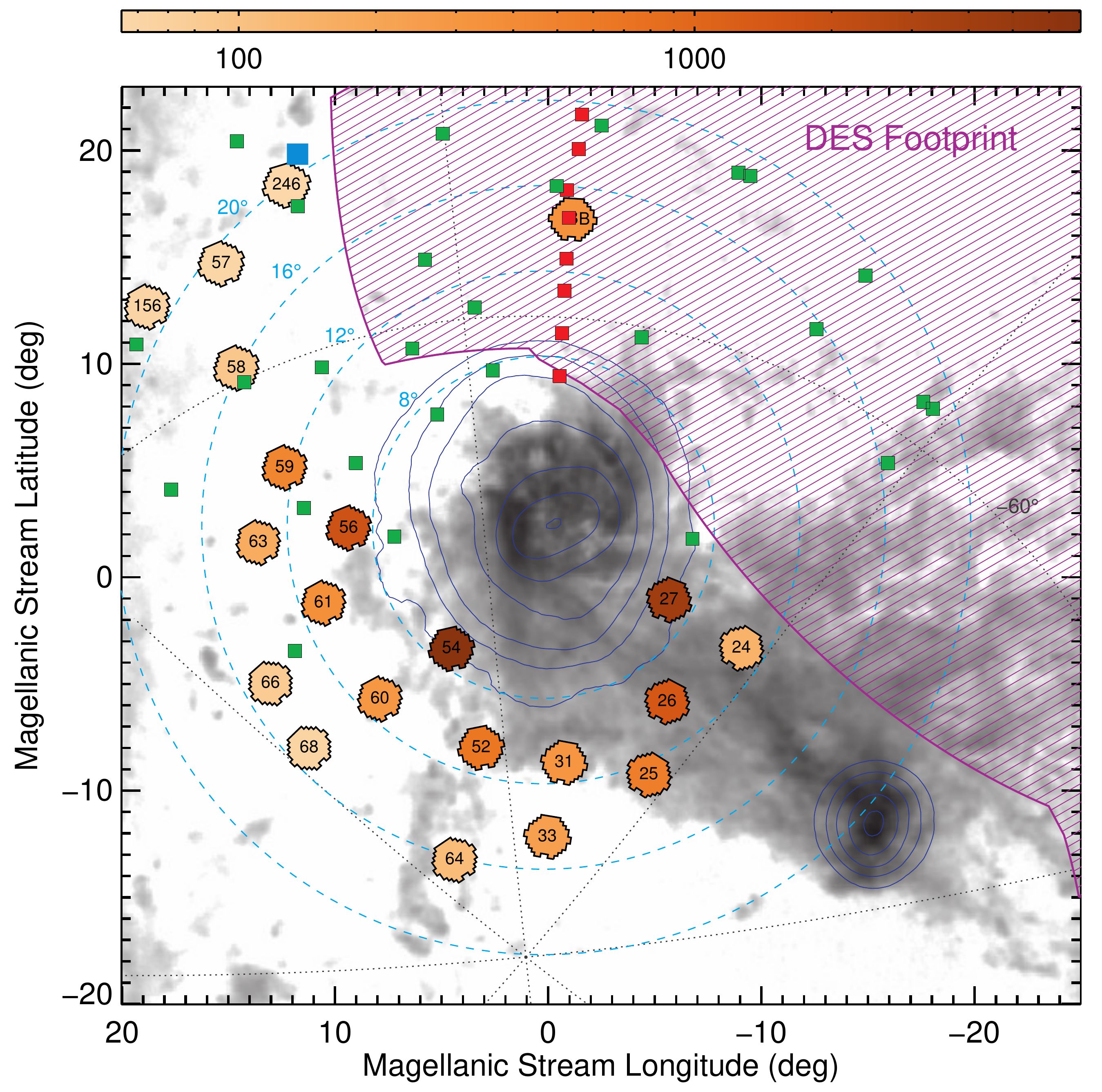}
\includegraphics[width=0.49\hsize,angle=0]{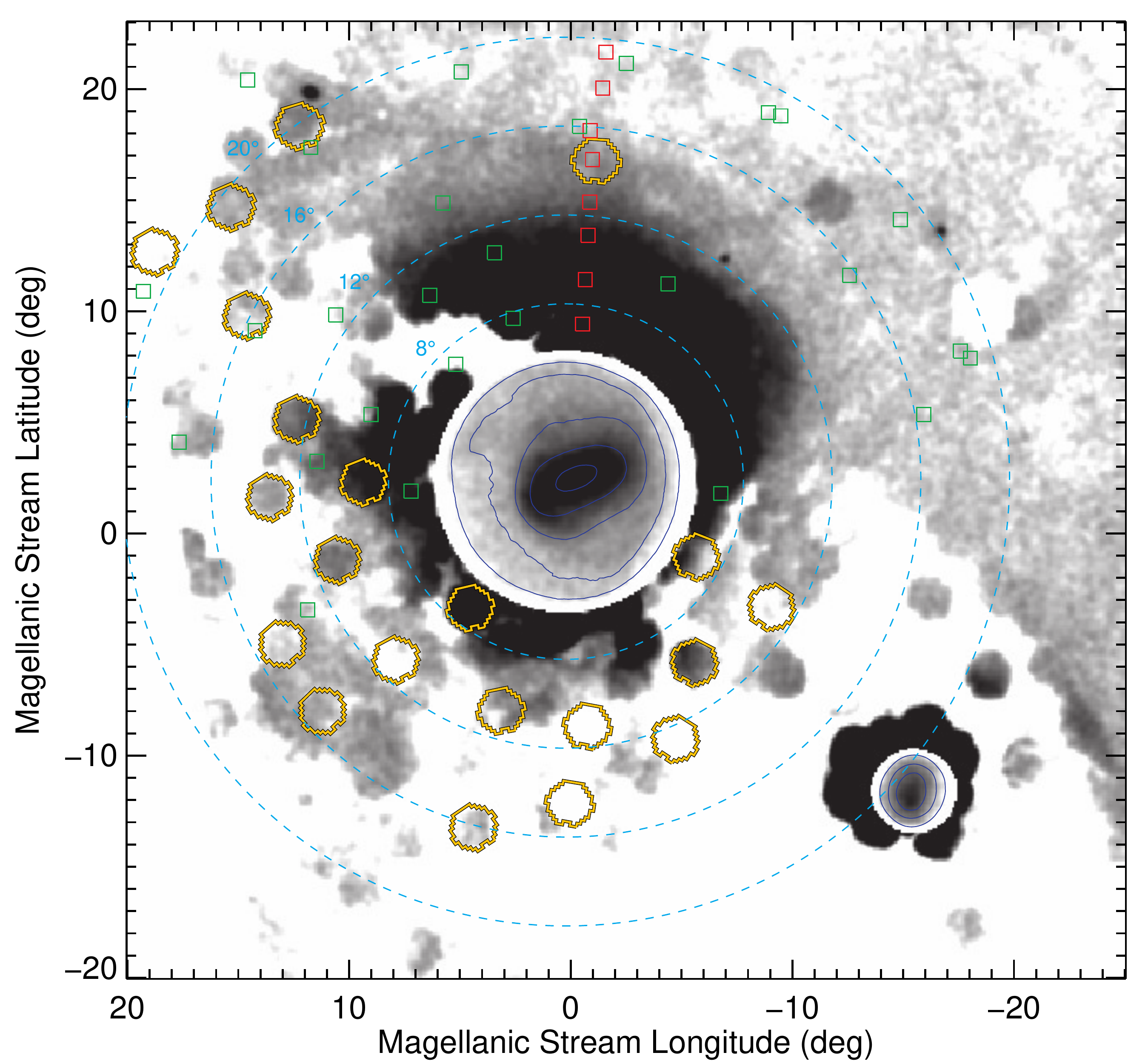}
\end{center}
\caption{The region of the Magellanic system relevant to the SMASH survey.  (Left) The observed \hi column density of the Magellanic Stream system is shown in grayscale \citep{Nidever2010}, while the dark blue contours represent the 2MASS \citep{Skrutskie06} RGB starcounts.  The 21 SMASH fields used in the current analysis are shown as filled orange hexagons with brightness indicating the LMC MSTO density in stars deg$^{-2}$ (see section \ref{sec:analysis}).
%, while other deep SMASH fields are indicated by open black hexagons.
The \citet{Majewski2009} fields are shown as filled green squares and \citet{Saha2010} as filled red squares.
The DES footprint is represented by the purple shaded region. (Right) LMC MSTO density map using the NOAO Source Catalog.  The central regions of the Magellanic
Clouds are represented by 2MASS RGB starcounts (grayscale and dark blue contours).
%\citep[NSC; ][]{Nidever2018}. Stars were selected with 0.0$\le$$g-r$$\le$0.4 and 21.8$\le$$g$$\le$22.8.
The irregular structure of the LMC disk with steep density dropoffs at $\sim$10--12\dgr (or $\sim$9 kpc at the LMC distance) is clearly visible.
The 21 SMASH fields are shown as orange hexagons, the MAPS fields as open green squares, and the OLS fields as open red squares.
%{\bf (Guy: "irregular" can be more specific)}
}
\label{fig_map}
\end{figure*}
% add Carina
% only show smaller region around LMC, and only fill in the fields we are discussing, leave rest open
% locations of newly discovered dwarfs

In the LMC periphery, \citet{Saha2010} used main sequence stars as tracers along a series of probes to the north of the LMC, and found LMC populations out to at least 16\dgr radius.  More recently, \citet{Mackey2016} discovered a spur-like structure in the northern LMC extending from $\sim$13\dgr to $\sim$16\dgr, pointing towards the Carina dwarf spheroidal galaxy.  \citet{Mackey2018} used new Dark Energy Camera \citep[DECam;][]{Flaugher2015} data in the southern and southwestern region of the LMC to map out the extended stellar populations in these regions using main sequence stars, and found a smooth LMC disk population extending to $\sim$10\dgr but with two spurs extending to $\sim$14\degr. One of these spurs is cospatial with the old RR Lyrae bridge discovered by \citep{Belokurov2016} and likely has a tidal origin, which is distinct from the Magellanic Bridge \citep[e.g.,][]{Noel2013,Carrera2017}. At still larger radius, \citet{Munoz2006b} discovered a kinematically ``cold'' population of high-velocity stars in the foreground of the Carina dwarf spheroidal galaxy that were consistent with having an LMC origin but $\sim$22\dgr from the LMC center.  The MAgellanic Periphery Survey (MAPS) was conducted as a follow-up of this discovery to ascertain the origin and structure of this new stellar population \citep{Majewski2009,Nidever2009}. It confirmed the existence of the extended stellar population and mapped it across an azimuthal range of over 180\degr.  \citet{BK16} used blue horizontal branch (BHB) stars to discover stream-like structures associated with the LMC out to distances of 30\dgr from the LMC center, indicating that the area of sky that may contain Magellanic Cloud populations is truly enormous.

The picture that emerges from these recent discoveries is one in which the disk of the LMC is both much more extended than previously thought, and, particularly the stellar populations
in the periphery, much more disturbed.  Clearly, the interaction between the Clouds \citep[e.g.,][]{Besla2007, Besla2012}
has left a strong imprint on the LMC's structure.  The debris and distortions from tidal interactions is a key observational
probe of the dynamical masses, orbits and interaction histories of the galaxies. 
However, a complete understanding of the origin of these substructures %found in the LMC disk and periphery 
requires a determination of the full extent and shape of the LMC stellar disk as well as other populations in the periphery. 

Here, we use deep SMASH photometry of old main-sequence turnoff (MSTO) stars to probe the spatial distribution and origin of stellar populations in the LMC periphery to lower surface brightnesses than previously possible.  We detect very extended stellar populations in many directions to $R$$\sim$21\dgr or 18.5 kpc (at the LMC distance)
with surface brightnesses as low as $\approx$34 mag arcsec$^{-2}$ in the $g$-band.  The layout of this paper is as follows.  Section \ref{sec:observations} gives an 
overview of our observations and data reduction.  Section \ref{sec:analysis} describes the analysis of our deep color magnitude diagrams and the results are presented in 
Section \ref{sec:results}.  The relevance and interpretation of our measurements are discussed in Section \ref{sec:discussion}, and, finally,
our conclusions are summarized in Section \ref{sec:summary}.

% Observations and Data Reduction
\section{Observations and Data Reduction}
\label{sec:observations}
The SMASH survey is a NOAO community survey that used $\sim$40 nights with the DECam on the CTIO Blanco 4m telescope to perform deep imaging in $ugriz$ of $\sim$200 ``island" fields spread over $\sim$2400 deg$^2$ of the southern sky, resulting in an extended map with $\sim$20\% filling factor.  In the inner regions of the Clouds, the observed fields fully overlap, yielding a filled map of the main bodies.  Figure \ref{fig_map} shows the region around the Magellanic Clouds and the SMASH fields including the 21 fields used in the current analysis.  Inner fields where the stellar populations are more complicated as well as outer fields with shallow data, heavily dust obscured or
with ambiguous LMC MSTO detection were excluded from our analysis of the LMC periphery.

The survey observations and data reduction are fully described in \citet{Nidever2017}.
In brief, we used the InstCal image data products (calibrated, single-frame images) produced by the DECam Community Pipeline \citep[CP;][]{Valdes2014}
and provided by the NOAO Science Archive Server\footnote{\url{https://www.portal-nvo.noao.edu}}. The photometric measurements were performed with the DAOPHOT \citep{Stetson1987,Stetson1994} suite-based PHOTRED\footnote{\url{https://github.com/dnidever/PHOTRED}} pipeline \citep{Nidever2017}. PHOTRED was used to perform WCS fitting, single-image PSF photometry (using ALLSTAR), forced PSF photometry of multiple images with a master source list created from a deep stack of all exposures (using ALLFRAME), aperture correction, photometric calibration, and dereddening.  The photometric reduction resulted in measurements with median 5$\sigma$ depths of 23.9, 24.8, 24.5, 24.2, and 23.5 mags in $ugriz$, respectively.

%{\color{red}  IDENTICAL TO HYDRA II PAPER.  NEEDS TO BE REWRITTEN.}
%Finally, star-like detections are separated from extended sources by enforcing a cut on the Sextractor stellar probability index ($\textrm{prob}>0.8$).
%{\color{red} UNTIL HERE.}

We also use the NOAO Source Catalog \citep[NSC;][]{Nidever2018} that contains aperture photometry of nearly all public DECam data including many SMASH 
exposures to select LMC MSTO stars and contiguously map the regions around the LMC (see Fig.\ \ref{fig_map})
with an eye to using both the SMASH and NSC data sets to further understand the structure of LMC.
The NSC selection is 0.0$\le$$g-r$$\le$0.4 and 21.8$\le$$g$$\le$22.8 with shape FWHM$<$1.5\arcsec~to remove background galaxies.
While the photometry was not dereddened, regions with SFD98 $E(B-V)$$>$0.5 mag were excluded.
%{\bf (Antonela: How were the background galaxies removed in this catalog?Because a careful removal of the background galaxies was done for SMASH. The NSC has the SMASH stars selected as described in this paper? or you used only the NSC to get the stars that were not in SMASH? and then you added the SMASH stars to the NSC to plot Fig. 1?)}

% from decam/plots/FieldB/
\begin{figure}
\begin{center}
\includegraphics[width=1.0\hsize,angle=0]{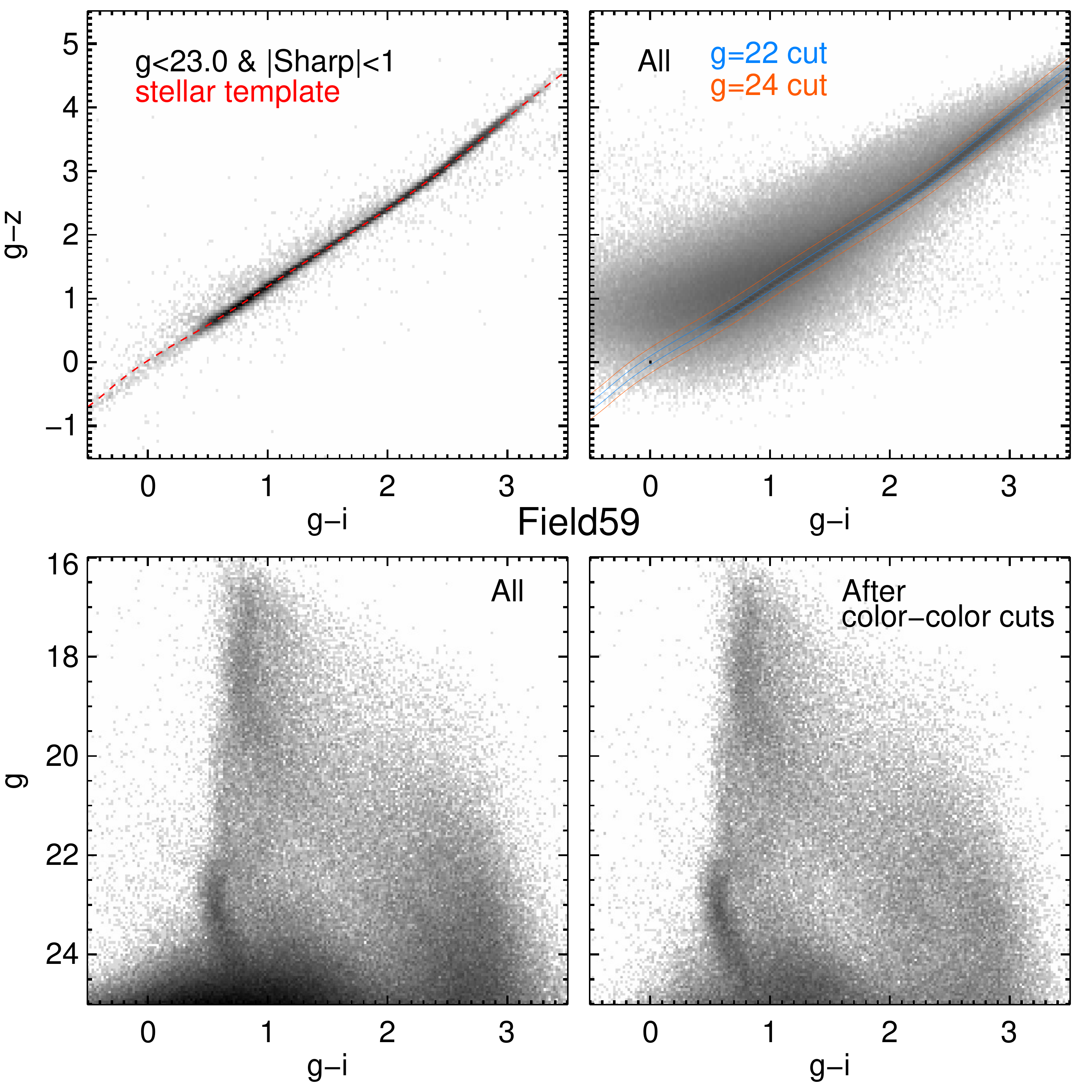}
\end{center}
\caption{Illustration of our star/galaxy separation using the stellar locus in one color-color space for Field59 (three color-color combinations are used for the final cuts).  Morphological shape cuts are already applied for the data in the figure. (Top left) $g-z$ vs.\ $g-i$ for $g$$<$23.0 and $|$Sharp$|$$<$1. The brighter sources are dominated by stars and are used to define the empirical template of the stellar locus (red dashed line).
(Top right) $g-z$ vs.\ $g-i$ for all stars.  The cuts around the stellar locus applied to objects at $g$=22 and $g$=24 mag are shown (blue and orange, respectively).
(Bottom left) $g$ vs.\ $g-i$ CMD for all
sources and (bottom right) sources that pass all color-color stellar locus cuts such as shown in the top panel. The large majority of extragalactic contaminants are removed.}
\label{fig_2cdcuts}
\end{figure}
% make grayscale
% show progression of cuts: raw CMD, CMD with morphological cuts, CMD with morph+stellar locus cuts
%    2CDs underneath

% make small modification
% overview, Nidever et al. (2016) in preparation

\section{Analysis}
\label{sec:analysis}

We conducted an analysis of the deep SMASH CMDs to measure the surface density of Magellanic stars using the MSTO.
The portion of the CMD populated with these stars is contaminated by both background galaxies and foreground
MW stars.  The galaxies were removed with both color-color and morphological cuts while MW models were used to subtract
the MW stars.  Then, the LMC MSTO luminosity function was compared to a fiducial field to measure the surface density.
The steps are described in more detail below.

%I think the section needs to start with a couple of sentences mentioning the various contaminants of pure MC populations that appear in the CMD, an ideally also the general idea on how they will be dealt with. Namely: galaxies (color color and morphological cuts, MW stars: MW models), etc. Then you describe the details. It would improve the readability to divide the section in subsections after this first introductory paragraph

\subsection{Removing Background Galaxies}

At faint magnitudes ($g$$\gtrsim$23), there is a large number of galaxies that contaminate our stellar signal of the LMC MSTO \citep[e.g.,][]{Fadely2012}.
Applying cuts on the morphological parameters only partially mitigates the problem because many of the galaxies are unresolved.
To further cull the stellar sample, we apply cuts taking advantage of the limited region of the multi-dimensional color space that stars occupy (i.e., the stellar
"locus") and the multi-band $ugriz$ SMASH photometry.  In each independent color-color diagram, bright PSF-like sources are used to determine the stellar locus
(and its intrinsic width) in that plane.  We use the $g-z$, $r-z$, and $i-z$ colors and $g-i$ as the fiducial color.
The $u$-band was not used since it is not as deep as the other bands, and colors using $u$ show higher intrinsic scatter (especially for bluer objects)
making it more challenging to remove galaxies
as ``outliers'' in these colors.  The stellar locus model is computed for each object (with good photometry in all three bands of a given
color-color plane) using its $g-i$ color.
%is interpolated to the $g-i$ value of each object with good photometry in all three bands of a given color-color plane.
If the deviation from the stellar locus is more than 2.5 times the observational color uncertainty or 0.2 mag then the object is considered
inconsistent with the stellar locus and removed.  A morphological shape cut is also applied using the DAOPHOT SHARP parameter \citep{Stetson1987}
which is a measure of the peakiness of an object's spatial profile compared to the PSF, with stars having values around zero while extended objects
have large values.  We set a lower cutoff of 0.2 on $|$SHARP$|$ for the brighter stars and allow the threshold to grow slightly
with magnitude but set an upper limit of 1.0.  An example of the stellar locus and morphology cuts is shown in Figure \ref{fig_2cdcuts}.

\subsection{Artificial Star Tests}
Artificial star tests (ASTs) were performed for all SMASH fields to estimate completeness.  In brief, artificial stars uniformly spanning the ($g-i$,$g$)
CMD space ($-$1.0$<$$g-i$$<$3.5, 17.0$<$$g$$<$27.0) were injected into all images for a single CCD and the images were processed with PHOTRED
the same way the original images were.  The final catalog of sources was then cross-matched with the original list of injected artificial stars.
See \citet{Choi2018a} for a more detailed description of the SMASH single--chip AST procedure.
%The star/galaxy separation algorithm was also applied to the ASTs to obtain the completeness for the SMASH star catalogs.
Completeness maps were generated for the CMD analysis using the requirement that an AST had to be detected in at least one image in both $g$ and $i$
and pass the star/galaxy separation cuts.
%in at least one $g$ image and at least one $i$ image.

\subsection{Cleaning Milky Way Populations}

MW populations are an important contaminant for our study of the MCs.  Popular MW models such as the Besan\c{c}on Galactic Model
\citep{Robin2003} or Galaxia \citep{Sharma2011} provide unsatisfactory results for our deep SMASH data that probe regions of the CMD not previously used to 
constrain the existing models.  Therefore, we fit our own MW model to the SMASH data itself.
%The approach is similar to the Besan\c{c}on model
Spatial and population parameters for the MW (see \citealt{Robin2003} for more details)
were constrained by fitting the observed distribution of stars in a set of CMDs.  To obtain the best constraint on the parameters of the different
MW components, as large an area as possible was included in the fit. SMASH fields were selected in the southern hemisphere far from the MW midplane
($\midline$b$\midline$$>$20 deg) and the MCs ($>$10 deg from each) to
avoid populations not included in the model. The set of 38 fields were fit in the ($g$,$g-i$) plane using Dartmouth isochrones\footnote{\url{http://stellar.dartmouth.edu/models/}} \citep{Dotter2008} in the region $-$0.2$<$$g-i$$<$1.8 and 14$<$$g$$<$22.5 to sample
an area that excludes faint M-dwarfs (which are challenging to model) and where the data are 100\% complete.
The resulting best-fit parameters for the thin disk, thick disk and halo are mostly consistent with previous studies.  Most
importantly to our present work, the best fit to the data gives a MW halo component that
follows a power law index of $-$3 and flattening $\sim$0.5 similar to those seen previously \citep{Bell2008,Deason2011,Slater2016}.
A more detailed description of this work will be presented in the near future (de Boer et al.\ 2018, in preparation).
MW models were generated for each SMASH field using the best-fit parameters, and convolved with photometric error and completeness of each individual field.

The SMASH stellar catalogs use \citet[hearafter SFD98]{sfd98} $E(B-V)$ dereddening by default.  However, CMDs of SFD98-dereddened photometry still 
showed slight field-to-field color offsets in the blue edge of the MW populations.  Therefore, we recomputed field-specific reddening using color
offsets of the MW blue edge.   The edge was measured in the magnitude range 21.0$<$$g$$<$22.0 for the MW model, the fiducial field, and the science
field.  The color offset relative to the MW model was then used as the $E(g-i)$ reddening and used to compute the $A(g)$ extinction (with \citealt{Schlafly2011}
extinction coefficients) for the fiducial and science data.

\begin{figure}
\begin{center}
\includegraphics[width=1.0\hsize,angle=0]{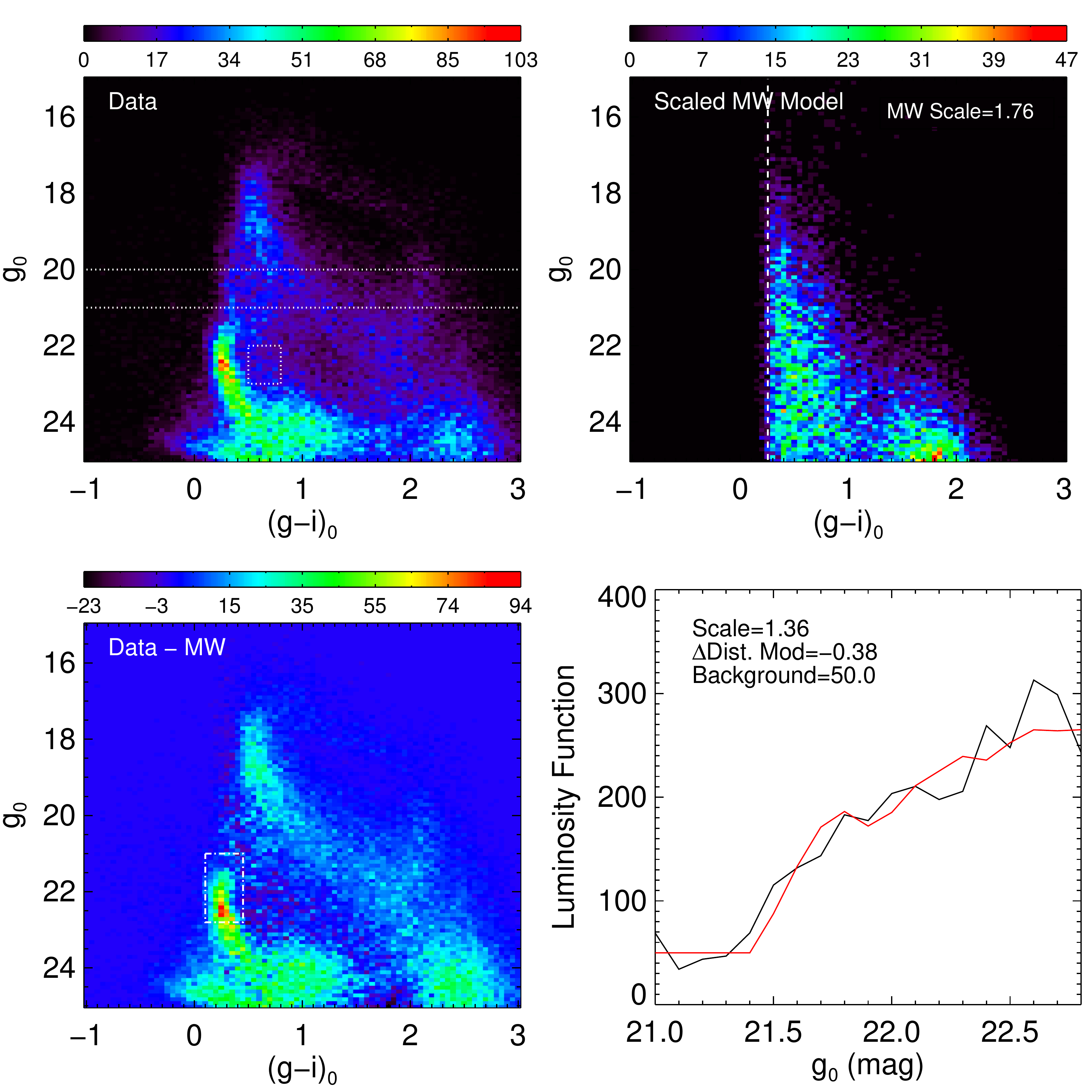}
\end{center}
\caption{The CMD analysis for a single SMASH field (Field59).  (Top left) Hess diagram of dereddened SMASH stars.  The magnitude range used to determine
the MW blue edge and the MW halo scaling box are shown in dotted lines.  (Top right) Scaled MW halo density map and the measured blue edge (dashed line). 
(Bottom left) Observed Hess diagrams with the scaled MW halo subtracted.  The box used to compute the luminosity function is shown in dashed-dotted lines.
(Bottom right) Observed LMC MSTO luminosity function (black) and the scaled, shifted and offset model using the fiducial field (red).}
\label{fig_cmdanalysis}
\end{figure}

Of the three MW populations (thin disk, thick disk, halo) only halo stars are present in the region of the CMD populated by the LMC MSTO.  The halo component
of the MW models was sometimes underrepresented (possibly because the faintest magnitudes were not included in the MW fitting)
and we used a scaling factor to match the data.  The region 0.50$\le$$(g-i)_0$$\le$0.80 and 22.0$\le$$g_0$$\le$22.5
(see Fig.\ \ref{fig_cmdanalysis}) was used to measure the observed and MW model densities and to calculate the MW halo scaling factor with a median value of 1.76.
The Hess density map in
the dereddened CMD was generated for the observed SMASH stars and then the scaled MW halo model population map (corrected for completeness) was subtracted.  It was
more challenging to remove the MW halo component in the fields nearer to the MW midplane because of the presence of the Monoceros ring stars in the foreground
(at $\sim$10--20 kpc).  For these fields the MW halo selection box was fine tuned for each field and sometimes the MW halo scaling parameter was
manually set after by-eye inspection of the residual image.

% LUMINOSITY FUNCTION FITTING

\begin{figure*}
\begin{center}
$\begin{array}{ccc}
\includegraphics[width=0.33\hsize,angle=0]{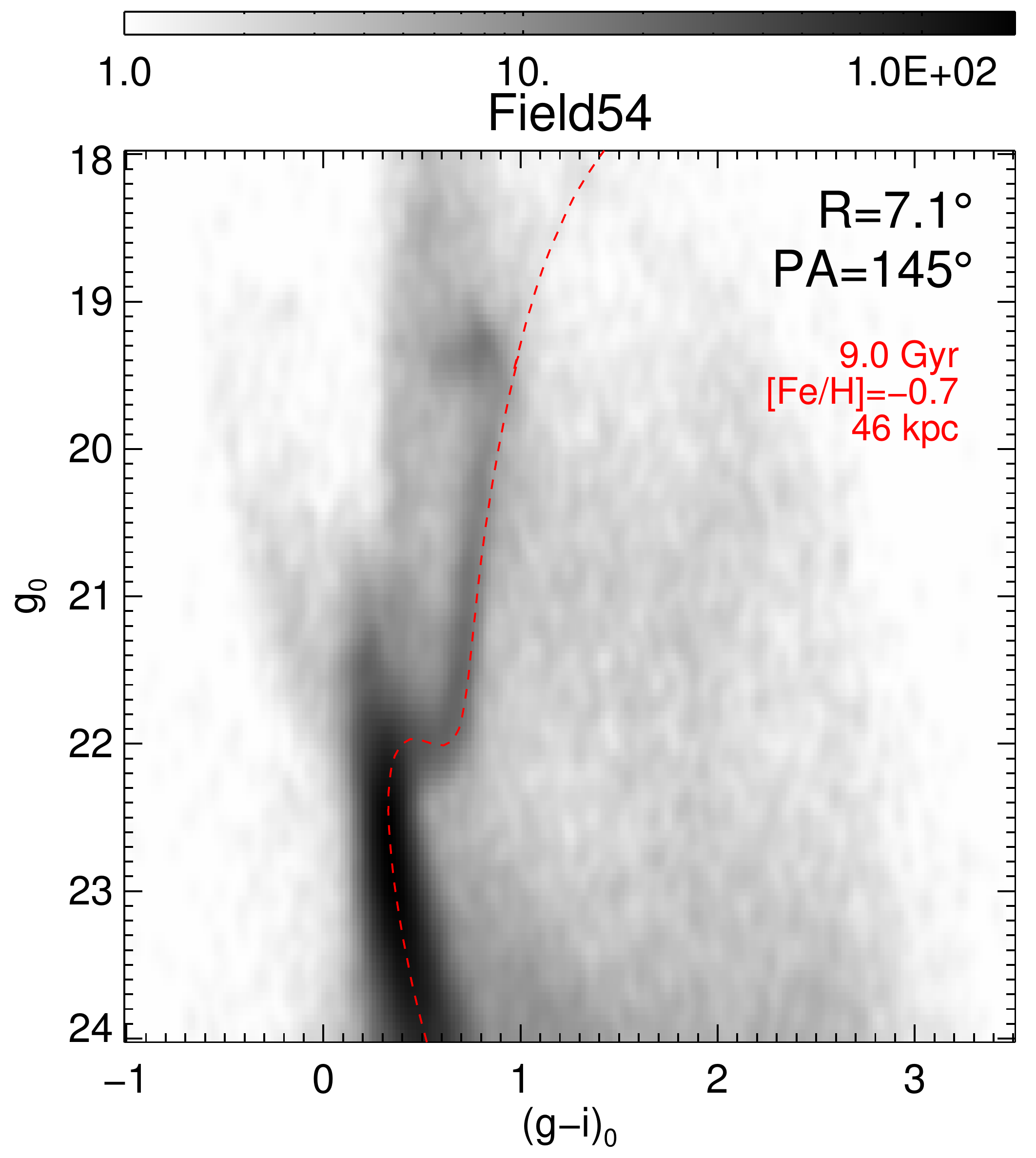}
\includegraphics[width=0.33\hsize,angle=0]{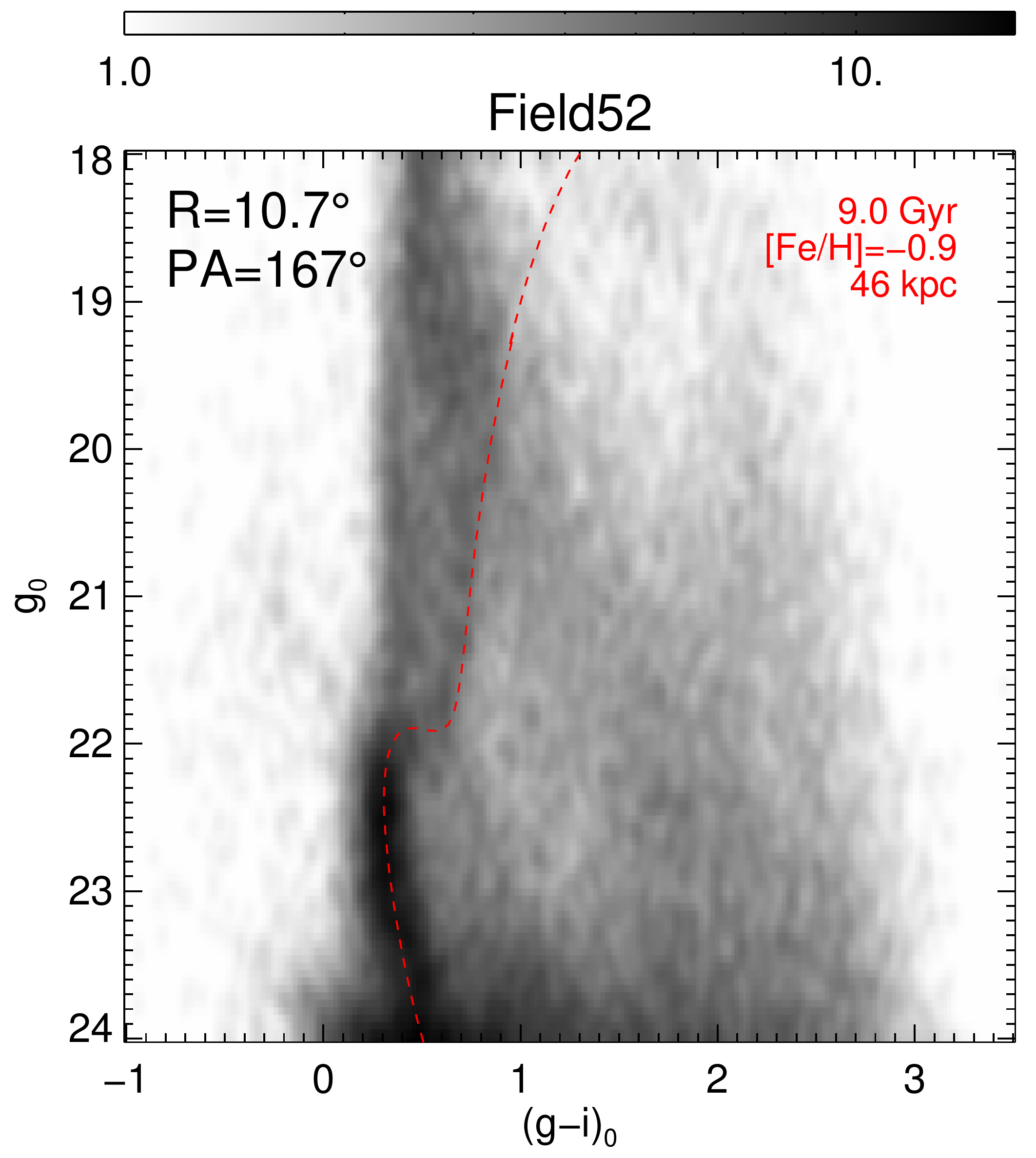}
\includegraphics[width=0.33\hsize,angle=0]{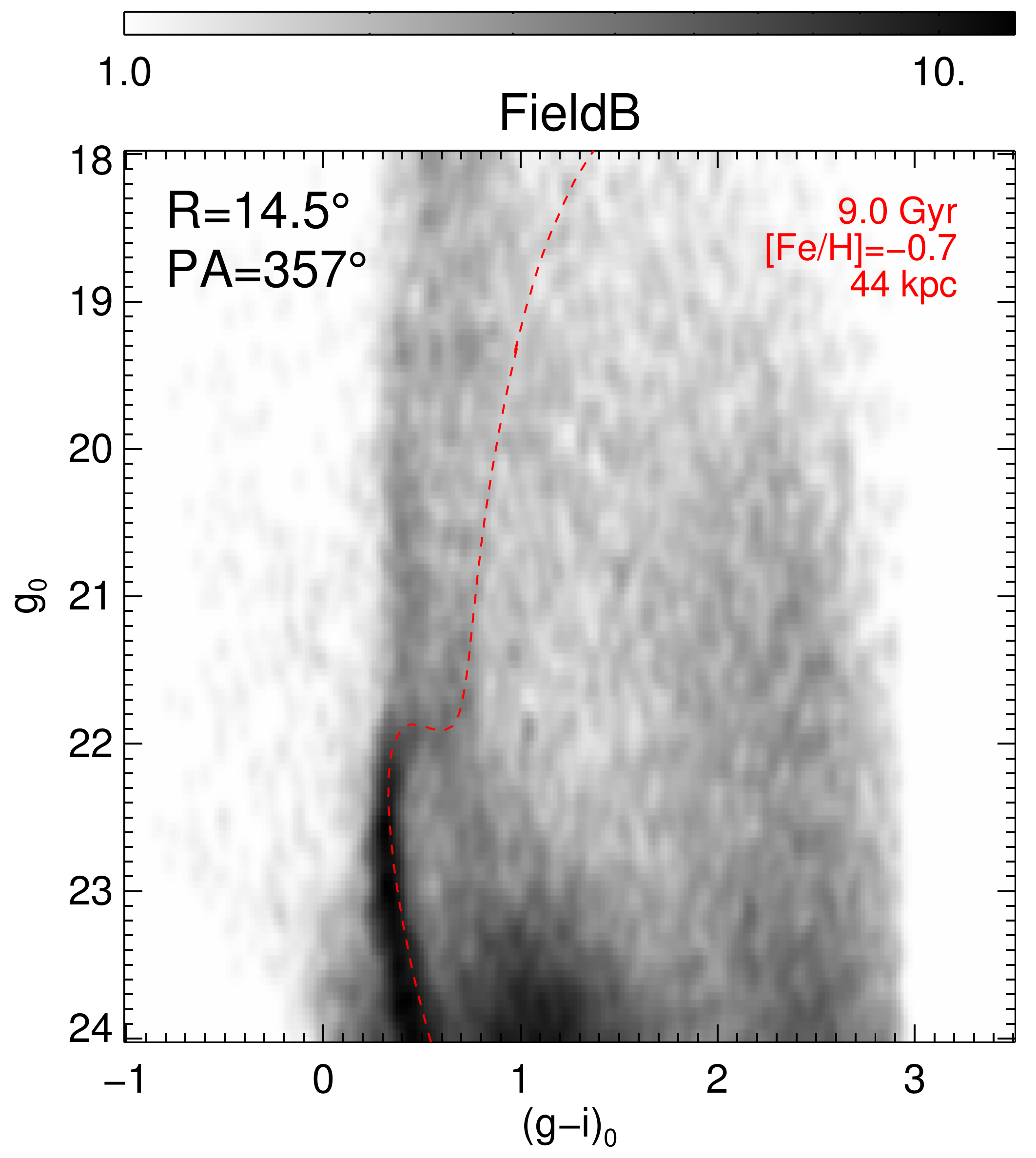} \\
\includegraphics[width=0.33\hsize,angle=0]{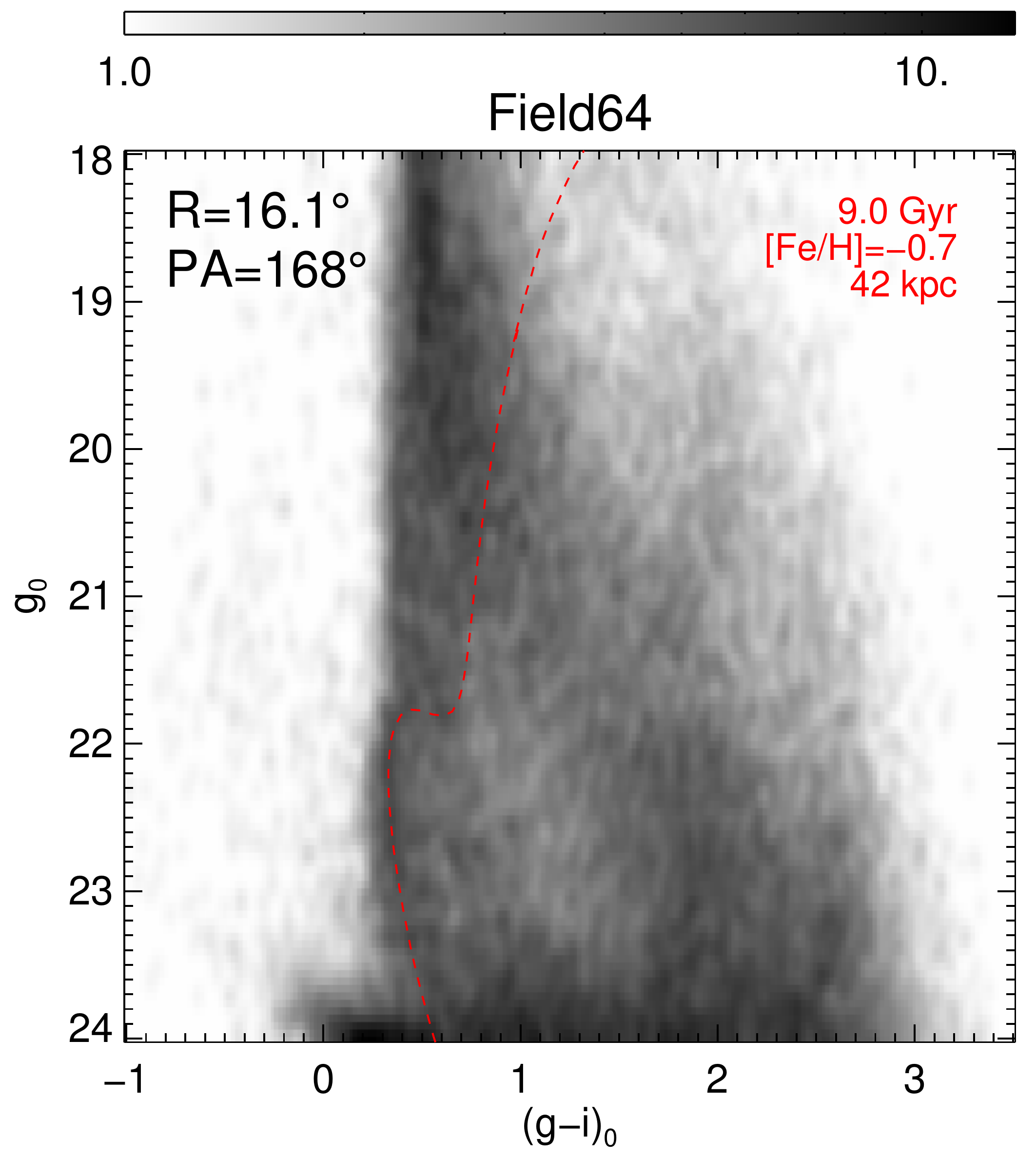}
\includegraphics[width=0.33\hsize,angle=0]{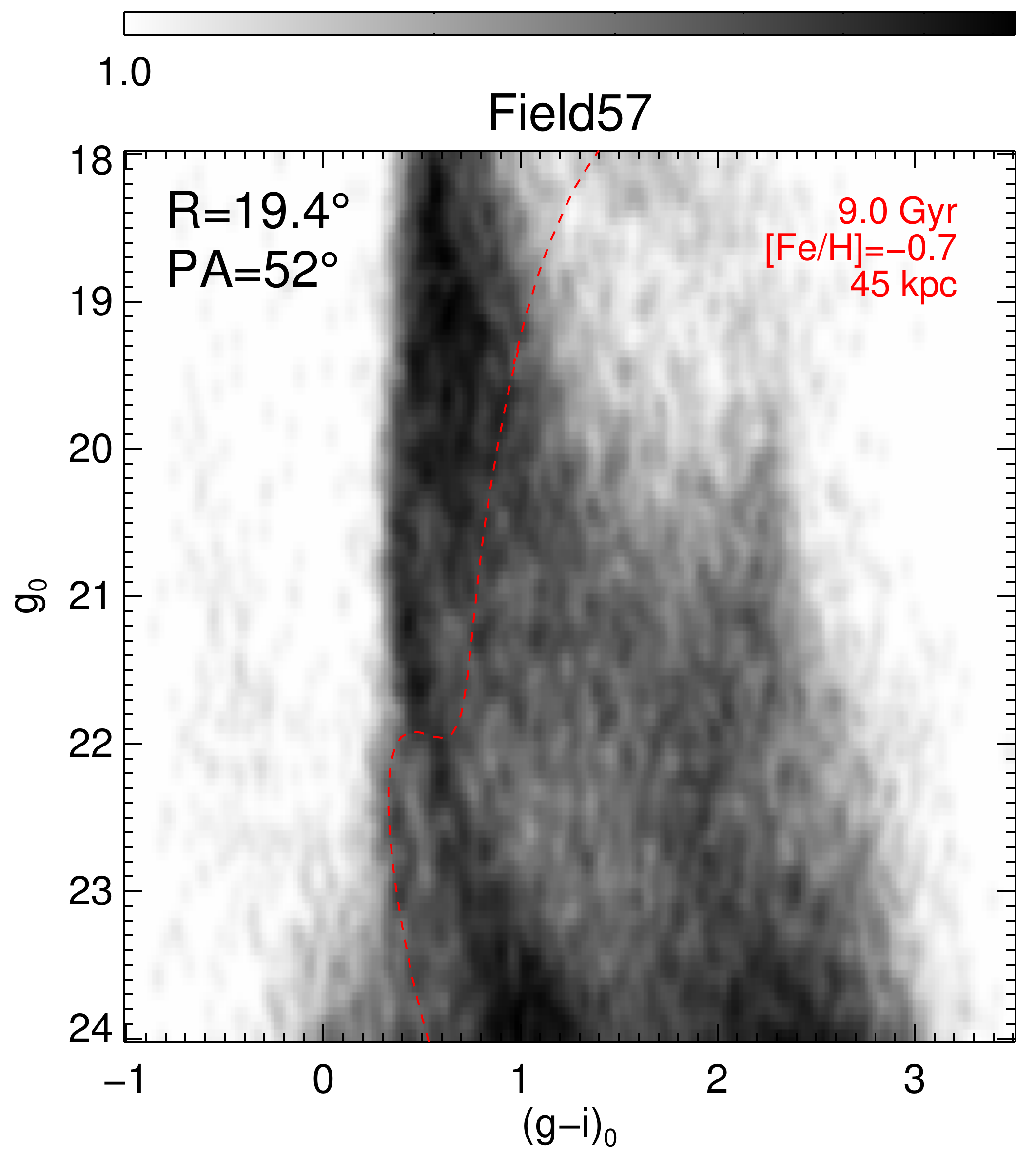}
\includegraphics[width=0.33\hsize,angle=0]{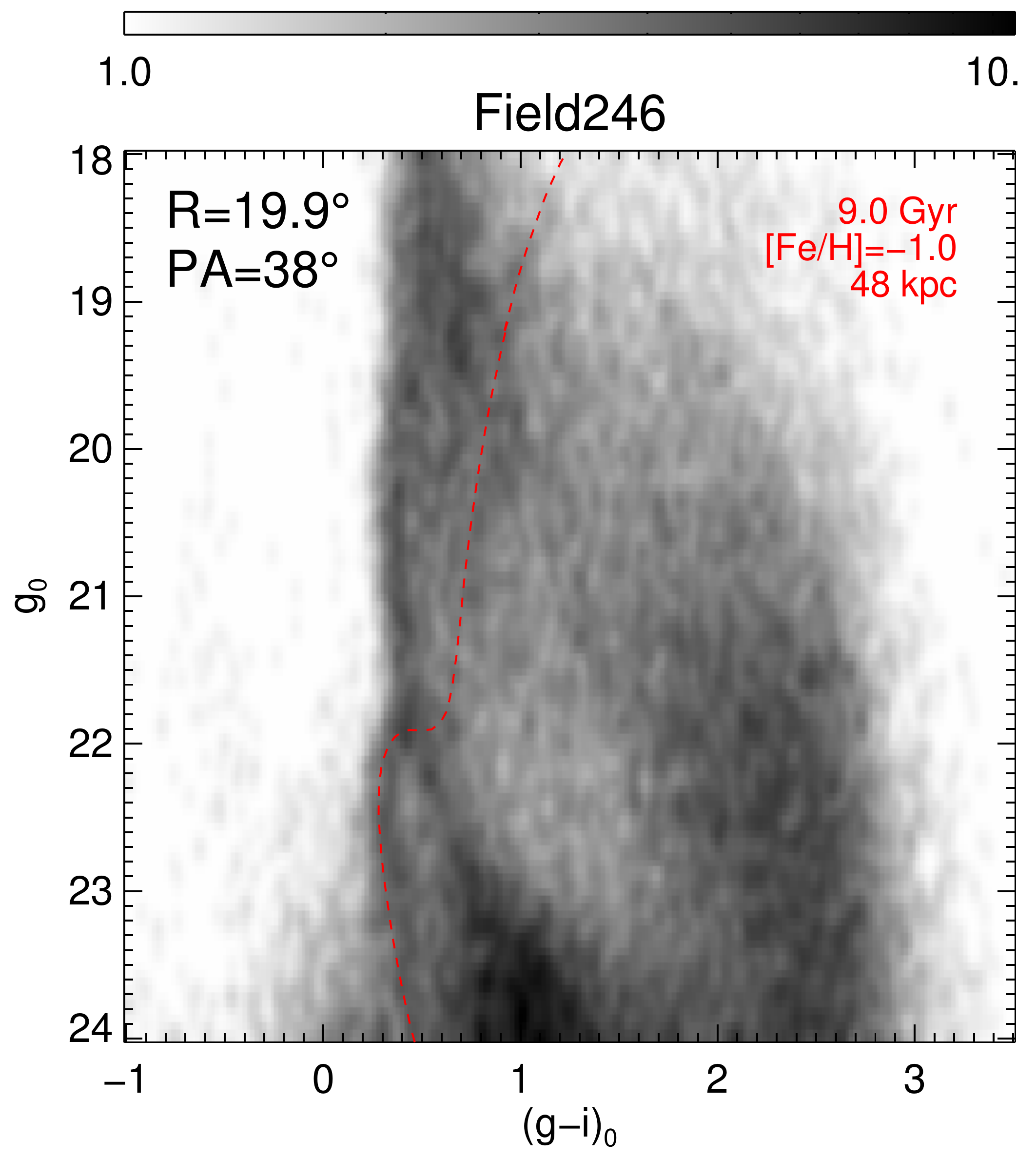}
\end{array}$
\end{center}
\caption{Color magnitude diagrams of six SMASH fields moving radially outwards from the center of the LMC: (Top left) Field 54, $R$=7.1\degr;
(Top middle) Field 52, $R$=10.7\degr; (Top right) Field B, $R$=14.5\degr; (Bottom left) Field 64, $R$=16.1\degr; (Bottom middle) Field57, $R$=19.4\degr;
and (Bottom right) Field 246, $R$=19.9\degr.  Best-fitting PARSEC isochrones \citep{Girardi2002} are shown in red with the parameters in the legend.
While only the innermost field (Field55) shows clear signs of young stars (and a double-subgiant
branch), faint main-sequence stars are seen in all fields to a radial distance of 21.1\degr or 18.4 kpc at the LMC distance.
}
\label{fig_cmds}
\end{figure*}

\subsection{Main-Sequence Luminosity Functions}

For our CMD analysis we use one SMASH field, FieldB north of the LMC (see Fig.\ \ref{fig_map}), with a well-measured LMC main-sequence and good
photometry with little extinction as our ``fiducial'' LMC field (see Fig.\ \ref{fig_cmds}).
%{\bf (Guy: How does FieldB vary with respect to the modeling done? It must deviate as a fiducial in some regions?)}
FieldB is at $R$=14.5\dgr and the data were obtained during the SMASH pilot project and is
therefore a magnitude deeper than the regular SMASH data.

\begin{figure*}
\begin{center}
\includegraphics[width=1.0\hsize,angle=0]{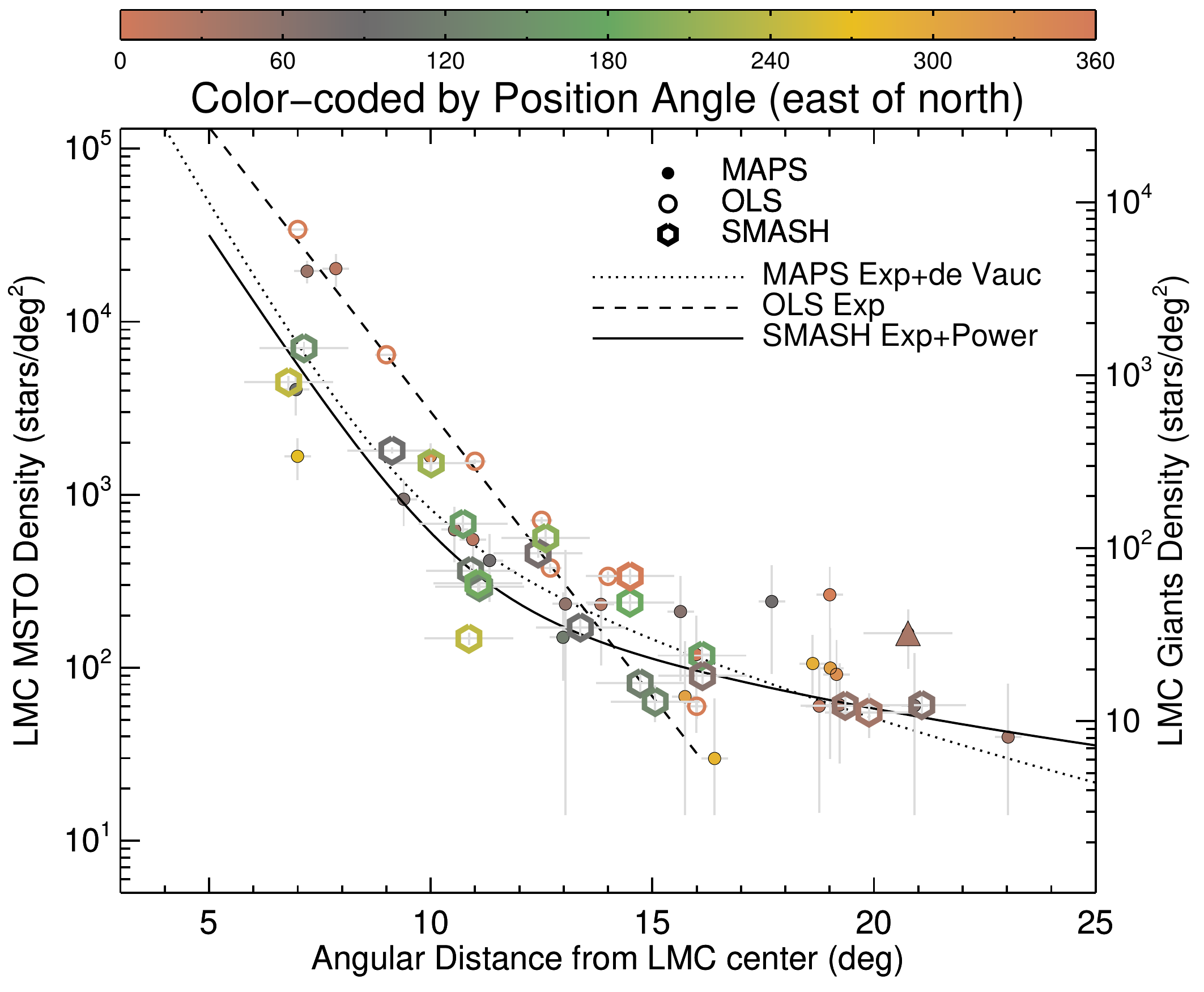}
\end{center}
\caption{The density of LMC stars versus radius for various studies.  The MAPS fields are represented by small filled dots, open circles
are data from \citet{Saha2010}, and open hexagons are the SMASH fields.  The group of LMC stars in the foreground of the Carina dwarf spheroidal discovered in \citet{Munoz2006b} is the large filled triangle.  The points are color-coded by position angle (East of North).  The SMASH 
The MAPS exponential plus de Vaucouleurs fit is shown as the dotted line, the OLS exponential fit is the dashed line, and the exponential plus power-law fit
is the solid line.}
\label{fig_density}
\end{figure*}

We used the LMC main-sequence luminosity function to measure the density of LMC stars relative to the fiducial field.
The region 0.11$\le$$(g-i)_0$$\le$0.44 and 21.0$\le$$g_0$$\le$22.8 (slightly fainter for the fiducial field) was used to measure
the observed luminosity function which extends somewhat brighter of the LMC MSTO to include a region to set the ``background'' level.  
The luminosity function was modeled using a shifted, scaled and offset version of the fiducial luminosity function (see Fig.\ \ref{fig_cmdanalysis}).  The
mean background level was $\sim$50 stars per 0.1 mag bin while the number of LMC MSTO stars per bin ranged from 25 to 3100.
This technique was found to be more reliable than a 2D modeling method which was more sensitive to the shape and position of the LMC main-sequence
and consistently underestimated the LMC densities.  The density of MSTO stars was then constructing by multiplying the scaling of the fiducial
luminosity function by the number of fidicual MSTO stars with $g$$\leq$22.8 (1014 stars) and dividing by the 3 deg$^2$ field-of-view of DECam
(see Fig.\ \ref{fig_density}).  The $g$-band surface brightness for the fiducial FieldB was calculated to be 32.02 mag arcsec$^{\-1}$ by comparing
the number of MSTO stars to synthetic photometry from a Z=0.002, 8 Gyr, 50 kpc BaSTI isochrone \citep{Pietrinferni04}.  Surface brightnesses for the other fields
were bootstrapped off of the FieldB value by using the luminosity function scaling value.  Table \ref{table_density} gives density and other information
on the 21 fields used in this study.

% DISCUSS BRIEFLY THE ISOCHRONE FITTING??

% Results
\section{Results}
\label{sec:results}

% what about the 24 and 29 deg fields? maybe there's stuff in the 24 deg field, could use 29 deg field for background (would be good to show a "background" field).
% would be good to analyze all the fields the same way and make similar plots for all.

% other interesting fields:
% 60, 153-155, 175?, 26 (for symmetry)
% Field59 has riz (20140120) and ug (20140126) catalogs, need to calibrate and merge them, but need g-r to calibrate g and r-mag

% we have reduced 0.9m data for Field55, Field64 and Field57

% show region in which we're making the number count, MSTO
%
% Fields 52, 64 and 156 are uncalibrated, do we have 0.9m data?
% or can I finished the jan2014 calibration for these fields

The right panel of Figure \ref{fig_map} shows the density map of LMC MSTO stars from the NSC. The features described in \citet{Mackey2016}
are visible such as the northern spur and its extension to the west as well as the sharp density dropoff on the western side of the LMC.
In addition, the newly discovered tidal spurs in the southern LMC by \citet{Mackey2018} are also partly visible in the NSC image.  Moreover,
the NSC data reveal previously unexplored structures on the eastern and northeastern side of the LMC,
although the photographic isophotes from \citet{Irwin1991} showed similar features but at smaller radii.
A fairly sharp edge exists on the western side (possibly giving rise to the large variations in the SMASH densities in this area)
that is reminiscent of the dropoff on the eastern side seen by Mackey et al. 
The northern spur and similar low-density features are extended farther
to the northeast (towards Carina) than compared to \citet{Mackey2016}.  Overall, the shape of the LMC at $\sim$11\dgr is quite striking and somewhat
triangular.
%even suggestive of an {\bf (Eric \& Antonela have no idea what this means)} ``annular sector''.

Figure \ref{fig_cmds} presents example Hess diagrams of objects passing the star/galaxy separation for six of the 21 SMASH fields used in this
study of the LMC periphery moving radially outward with the best ``by-eye'' fit isochrones.
%some example Hess diagrams of six (out of 21) SMASH fields in the periphery of the LMC moving radially outward with the best by-eye fit isochrones. 
The LMC lower main-sequence is clearly visible in all fields indicating that the deep SMASH data can
be used to trace Magellanic populations to very low surface brightnesses.

Figure \ref{fig_density} shows the radial density profile of the LMC from individual SMASH (open hexagons), MAPS (small filled circles) and OLS (open circles) fields.
Each symbol is color-coded by its position angle.  Both the SMASH and OLS densities were put on the MAPS scale of giants stars per deg$^2$ using
overlapping fields that have well-measured values.  While the three datasets cover slightly different regions of the sky their results generally
agree with each other.  All of them show an exponential decline out to $\sim$15\dgr with a scale-length of $\sim$1\degr.
%although the SMASH results exhibit significant scatter in the $\sim$11\dgr fields which might suggest the existence of substructure.
Both the MAPS and SMASH data show the existence of low-density populations
beyond $\sim$16\dgr extending to $\sim$21\dgr with the SMASH best-fit inner exponential ($h_{\rm R}$=1.1$\pm$0.03\degr) and outer power-law
($\alpha$=$-$2.2$\pm$0.4) exhibiting a ``break'' at $\sim$13--15\degr.
The SMASH results show significant scatter of $\sim$70\% around the the best-fit model, especially as a function of position angle, with some fields
deviating by as much as 2--2.5$\times$ from the best-fit model.
At smaller radii this is due to the inclined and intrinsically elongated LMC disk \citep{vandermarel2001} but at larger radii this is likely because
of the irregular shape of the LMC (as seen by Mackey et al.\ and the NSC data) as well as the substructure of spurs and stream-like features.
The total mass of the power-low component (from 13--23\degr), assuming it continues on the northwestern side not probed by SMASH, compared to the
exponential component is 0.4\%, although this quantity is quite uncertain due to the large variability in the LMC envelope.

% Discussion
\section{Discussion}
\label{sec:discussion}

The origins of the stellar populations in the LMC periphery remain unclear.  \citet{Majewski2009} proposed that the large spatial extent, radial profile,
radial velocity profile, and low metallicity resembled a classical, accreted stellar halo.  In contrast, \citet{Saha2010} found that out to $\sim$16\dgr the radial
profile is well-fit by an exponential and, therefore, the stellar populations are likely an extended disk.  Morever, \citet{Mackey2016}
claimed that the northern stream-like feature was material most likely stripped from the outskirts of the LMC disk by the MW's tidal force.
\citet{Besla2016}, \citet{Choi2018a} and \citet{Choi2018b} discuss substructures in the LMC disk that likely arose due to the recent interactions
with the SMC, and, potentially, these impulses could have created similar structures at larger radii.
% streams?

The SMASH data reveal a very extended ($R$$\sim$21\degr) ``envelope'' of old ($\sim$9 Gyr) and relatively metal-poor ([Fe/H]$\approx$$-$0.8 dex) stars
around the LMC covering a large range in position angle and possessing a shallow radial density profile at large radii that constitutes $\sim$0.4\%
of the LMC's stellar mass.
Our results corroborate earlier work on the existence of this structure but add important empirical constraints using our deep CMDs
especially in the south, east and northeastern side of the LMC periphery.
Our NSC map extends contiguous mapping to the eastern side of the LMC and uncovers an irregular shape and sharp density dropoff there ($\sim$10\degr).
This, combined with the Mackey et al.\ results, illustrates the pronounced asymmetric shape of the LMC at $\sim$10\degr.
%{\bf (A sentence suggested by Guy: However, the origin of this structure remains unclear and none of the specific attributes previously assigned (halo, disk, stripped stream, etc.) can be affirmed with the present data. We prefer to call this extended structure a  faint LMC envelope.)}
In addition, the SMASH densities exhibit large scatter around our best-fit exponential plus power law model of $\sim$70\%.  
All of this shows that the structure of the LMC stellar periphery is highly disturbed and irregular quite possibly due to  
tidal interactions with the SMC and MW.
%is surrounded by an envelope of  disturbed old stars likely perturbed by tidal interactions with the SMC and MW.
%with many regions having quite thin main sequences suggestive of small spreads in age and distance. 
%The shape of the LMC at $\sim$10--12\dgr as seen by Mackey et al.\ and our NSC data is very irregular
%with sharp dropoffs in density on the eastern and western sides.

The recent discovery of several dwarf satellite galaxies \citep[e.g.,][]{Bechtol2015,Drlica-Wagner2015} many of them close to the MCs and
some having radial velocities consistent with MC origins \citep[e.g.,][]{Walker2016,Li2018} indicates that the MCs
have their own system of satellites as suggested by simulations \citep{Deason2015,Wheeler2015}.  It is therefore quite likely that the MCs
also have a classical halo of disrupted satellite galaxies.
We also note that the mass fraction that we derive for the LMC stellar envelope of $\sim$0.4\% is quite similar to the mass fraction of the
MW's stellar halo of $\sim$1\% \citep{Carney1990,Bell2008} and expectations from simulations of stellar halos \citep[e.g.,][]{Bullock2005}.
In addition, RR Lyrae studies have found that the inner region of the LMC possesses a kinematically-hot and metal-poor
([Fe/H]$\approx$-1.5 dex) stellar component \citep{Minniti03,Borissova2004}.  Furthermore, recent SDSS-IV APOGEE-2 \citep{Majewski2017,Blanton2017}
spectroscopic results also show the existence
of a kinematically-hot and metal-poor (to [Fe/H]$\approx$$-$2.5 dex) population of RGB stars prominent at intermediate radii
($\sim$7--9\degr; Nidever et al.\ 2018, in preparation).  Therefore, there is strong evidence from a variety of studies suggesting
that a classical accreted halo of the LMC does exist and is roughly consistent with our observations of the LMC periphery.

On the other hand, the Magellanic interaction simulations of \citet{Besla2016} and \citet{Mackey2016} compellingly reproduce several
substructures in the outer LMC.
% Tidal induced perturbations can produce N/S asymmetries explaining the observed higher densities in the north
Tidally induced perturbations in the disk also create asymmetries and different density profiles in the north versus the
south that can explain the observed higher densities in the northern portion of the LMC periphery.
% outer disk known to be old and metal-poor
Moreover, the outer LMC disk is known to be old and somewhat more metal-poor compared to the
inner regions \citep{Carrera2008,Meschin2014} which could fit with the known characteristics of the periphery stars.
% Simulations show dwarfs have extended, halo-like component
This trend is also seen in simulations of dwarf galaxy formation that suggest they should have an extended distribution of old, metal-poor stars
that resembles a ``halo-like'' component due to early star formation taking place in very low $V/\sigma$ gas \citep{Read2006,ElBadry2018}.
% LMC's disk is thick, so disk/halo distinction less clear
%We should also remember that
In addition, the LMC disk is substantially thicker than those of larger spirals with a $V/\sigma$=3 \citep{vanderMarel2002}
which is even lower than the MW's thick disk ($V/\sigma$=4) making the distinction between a disk and halo component less clear.

% Both disturbed disk and accretion contribute to stellar envelope.
Therefore, it is quite likely that both a tidally disturbed disk (with a potentially old, metal-poor extended component)
as well as the accretion of satellites (e.g., producing a classical halo)
%as a classical, accreted halo
contribute to the stellar envelope that we reveal here, 
but further information (e.g., stellar kinematics and abundances) will be needed to fully unravel their relative contributions.
% GURTINA: The LMC disk is clearly asymmetric, and likely more extended in the north than the south. it's worth mentioning this in the interpretation of Figure 5  - it's possible that the fields near carina would have a higher density than at similar angular radius in the south because of this (e.g. extended disk + existing outer halo population). This is related to the last paragraph of the discussion section where you bring up perturbations --- such perturbations induce asymmetries that can generate different density profiles in the north vs the south. 

\section{Summary}
\label{sec:summary}

We have used deep SMASH photometry obtained with CTIO-4m Blanco and DECam to study the low surface brightness features in the periphery of
the Large Magellanic Cloud.  Our data reveal a very extended stellar envelope around the LMC reaching to large radii with a distorted structure.
%and with substantial azimuthal variations in the brightness profile.
%a large stellar envelope of material surrounding the Large Cloud.
%We have used deep photometry to reveal a very extended stellar envelope around the LMC reaching to large radii and with substantial azimuthal
%variations in the brightness profile.  These results provide quantitative constraints on simulations of the formation of dwarf galaxies
%as well as the interaction of the Magellanic Clouds that will improve our understanding of these important systems.
Our main conclusions are:
\begin{enumerate}
\item We detect faint LMC main-sequence populations to very large radii in many directions reaching $R$=21.1\dgr in the northeast with a
surface brightness of $\Sigma_{g}$$\approx$34 mag arcsec$^2$.
\item The deep SMASH CMDs of the outer LMC reveal the stellar populations to be old ($\sim$9 Gyr) and relatively metal-poor ([Fe/H]$\approx$$-$0.8 dex).
\item The LMC surface density profile initially follows an exponential decline ($h_{\rm R}$=1.1$\pm$0.03\degr) but ``breaks'' to a shallower
slope at $\sim$13--15\dgr with a power-law index of $\alpha$=$-$2.2$\pm$0.4. The low-density outer component (13$\leq$$R$$\leq$23\degr) contributes $\sim$0.4\%
of the total stellar mass of the LMC.
\item The SMASH main-sequence densities show large azimuthal variations with a scatter of $\sim$70\% around the the best-fit model, but with
some deviating by 2--2.5$\times$ from the average density.  This shows that the structure of the outer LMC populations is highly distorted.
\item Using LMC main-sequence turnoff stars from the NOAO Source Catalog, we find a steep dropoff in density on the eastern side at $R$$\approx$8\degr,
similar to the dropoff seen by \citet{Mackey2016} on the western side, as well as a low-density feature extending far to the northeast.
\end{enumerate}

The origin of the outer LMC stellar envelope remains unclear with evidence for both a classical accreted halo and tidal stripping of the outer disk.
Additional information such as accurate distances and line-of-sight depths, kinematics and chemical abundances will be useful in unraveling the
mechanisms responsible for creating this extensive structure.
Finally, the low surface brightness measurements presented here should provide quantitative constraints on simulations of the formation of dwarf galaxies
as well as the interaction of the Magellanic Clouds that will help improve our understanding of these important systems.

\acknowledgments
Y.C. \& E.F.B. acknowledge support from NSF grant AST 1008342. TdB acknowledges financial support from the ERC under Grant Agreement n. 308024.
M-RC acknowledges support from the European Research Council (ERC) consolidator grant programme No. 682115.
RRM acknowledges partial support from CONICYT Anillo project ACT-1122 and project BASAL PFB-$06$. GSS is supported by grants from NASA.

Based on observations at Cerro Tololo Inter-American Observatory, National Optical Astronomy Observatory (NOAO Prop. ID: 2013A-0411 and 2013B-0440; PI: Nidever),
which is operated by the Association of Universities for Research in Astronomy (AURA) under a cooperative agreement with the National Science Foundation. This project used data obtained with the Dark Energy Camera (DECam), which was constructed by the Dark Energy Survey (DES) collaborating institutions: Argonne National Lab, University of California Santa Cruz, University of Cambridge, Centro de Investigaciones Energeticas, Medioambientales y Tecnologicas-Madrid, University of Chicago, University College London, DES-Brazil consortium, University of Edinburgh, ETH-Zurich, University of Illinois at Urbana-Champaign, Institut de Ciencies de l'Espai, Institut de Fisica d'Altes Energies, Lawrence Berkeley National Lab, Ludwig-Maximilians Universit\"at, University of Michigan, National Optical Astronomy Observatory, University of Nottingham, Ohio State University, University of Pennsylvania, University of Portsmouth, SLAC National Lab, Stanford University, University of Sussex, and Texas A\&M University. Funding for DES, including DECam, has been provided by the U.S. Department of Energy, National Science Foundation, Ministry of Education and Science (Spain), Science and Technology Facilities Council (UK), Higher Education Funding Council (England), National Center for Supercomputing Applications, Kavli Institute for Cosmological Physics, Financiadora de Estudos e Projetos, Funda\c{c}\~ao Carlos Chagas Filho de Amparo a Pesquisa, Conselho Nacional de Desenvolvimento Cient?fico e Tecnol\'ogico and the Minist\'erio da Ci\^encia e Tecnologia (Brazil), the German Research Foundation-sponsored cluster of excellence "Origin and Structure of the Universe" and the DES collaborating institutions.

%\bibliography{/Users/martin/Work/Papers/Biblio}
%\bibliographystyle{apj}

\begin{deluxetable*}{lccccccc}
\tablecaption{SMASH LMC MSTO Densities}
\tablecolumns{10}
\tablewidth{0pt}
\tablehead{
\colhead{Field Name\tablenotemark{a}} & \colhead{RA} & \colhead{DEC} & 
\colhead{$R_{\rm LMC}$} & \colhead{PA$_{\rm LMC}$} & 
\colhead{$\rho_{\rm MSTO}$} & \colhead{$\rho_{\rm MSTO}$ Error} & \colhead{$\Sigma_{\rm g}$} \\
\colhead{} & \colhead{(J2000)} & \colhead{(J2000)} & 
\colhead{(deg)} & \colhead{(deg)} &
\colhead{stars deg$^{-2}$} & \colhead{stars deg$^{-2}$} & \colhead{mag arcsec$^{-2}$}
}
\startdata
Field24 & 03:14:56.4 & $-$72:29:06.4 & 10.86 & 240.88 & 147.52 & 11.94 & 32.90 \\
Field25 & 03:21:03.6 & $-$79:48:24.7 & 12.59 & 205.19 & 562.89 & 87.06 & 31.45 \\
Field26 & 03:40:18.9 & $-$76:25:04.3 & 10.01 & 217.56 & 1521.46 & 119.00 & 30.37 \\
Field27 & 04:08:11.9 & $-$72:02:02.9 & 6.79 & 242.28 & 4475.79 & 391.49 & 29.20 \\
Field31 & 04:52:55.9 & $-$80:44:31.5 & 11.07 & 187.26 & 306.73 & 34.94 & 32.11 \\
Field33 & 04:57:23.1 & $-$84:17:33.7 & 14.49 & 182.99 & 238.26 & 26.17 & 32.38 \\
FieldB & 05:22:23.9 & $-$55:24:00.0 & 14.50 & 357.00 & 338.00 & 33.00 & 32.00 \\
Field52 & 06:25:26.4 & $-$79:59:53.0 & 10.73 & 166.53 & 678.80 & 73.38 & 31.24 \\
Field54 & 06:32:06.6 & $-$75:10:37.7 & 7.14 & 145.14 & 7044.11 & 601.96 & 28.70 \\
Field246 & 06:47:22.0 & $-$52:13:49.0 & 19.89 & 37.88 & 55.21 & 16.01 & 33.97 \\
Field56 & 07:09:11.3 & $-$68:18:59.1 & 9.13 & 92.34 & 1800.79 & 76.76 & 30.18 \\
Field57 & 07:14:33.6 & $-$54:37:25.5 & 19.35 & 51.83 & 60.38 & 10.54 & 33.87 \\
Field58 & 07:25:02.7 & $-$59:20:37.4 & 16.13 & 64.15 & 89.76 & 16.10 & 33.44 \\
Field59 & 07:25:10.3 & $-$64:31:15.1 & 12.42 & 78.96 & 458.57 & 28.28 & 31.67 \\
Field60 & 07:36:20.4 & $-$76:11:39.5 & 11.10 & 138.69 & 293.20 & 20.17 & 32.15 \\
Field61 & 07:38:41.3 & $-$70:59:30.6 & 10.90 & 111.18 & 363.04 & 29.46 & 31.92 \\
Field156 & 07:41:42.3 & $-$54:53:23.6 & 21.08 & 62.04 & 60.93 & 10.10 & 33.86 \\
Field63 & 07:52:11.1 & $-$67:00:16.0 & 13.38 & 95.13 & 171.09 & 21.21 & 32.74 \\
Field64 & 08:00:24.1 & $-$84:26:13.8 & 16.12 & 167.53 & 117.54 & 17.32 & 33.15 \\
Field66 & 08:32:37.3 & $-$72:39:13.8 & 14.73 & 122.08 & 81.37 & 11.18 & 33.55 \\
Field68 & 08:43:39.0 & $-$76:05:41.0 & 15.06 & 135.74 & 63.62 & 15.29 & 33.81
\enddata
\tablenotetext{a}{The MSTO densities are calculated using the fiducial scaling
factor multiplied by the number of MSTO stars with $g$$\leq$=22.8 (1014) divided by
the DECam field-of-view (3 deg$^2$)}
\label{table_density}
\end{deluxetable*}

% Bibtex will create a .bbs file in the directory and before sending to the editor, I should replace the bibliography call by this file.

\end{document}